\documentclass[aip,apl,reprint]{revtex4-1}
\usepackage{mathtools}
\usepackage{float}
\usepackage{amsfonts}
\usepackage{enumitem}
\usepackage{graphicx}
\usepackage{color}

\begin{document}
\title{Quantum-Enhanced Noise Radar}
\author{C.W. Sandbo Chang}
\affiliation{Institute for Quantum Computing and Electrical and Computer Engineering, University of Waterloo, Waterloo, Canada}
\author{A.M. Vadiraj}
\affiliation{Institute for Quantum Computing and Electrical and Computer Engineering, University of Waterloo, Waterloo, Canada}
\author{J. Bourassa}
\affiliation{Institut quantique and DŽpartement de Physique, Universit\'e de Sherbrooke, Sherbrooke, Canada}
\author{B. Balaji}
\affiliation{Radar Sensing and Exploitation Section, Defence R\&D Canada, Ottawa Research Center, ON, Canada K1A 0Z4}
\author{C.M. Wilson}
\affiliation{Institute for Quantum Computing and Electrical and Computer Engineering, University of Waterloo, Waterloo, Canada}
\email{chris.wilson@uwaterloo.ca}

\begin{abstract}
We propose a novel protocol for quantum illumination: a quantum-enhanced noise radar.  A two-mode squeezed state, which exhibits continuous-variable entanglement between so-called signal and idler beams, is used as input to the radar system.  Compared to existing proposals for quantum illumination, our protocol does not require joint measurement of the signal and idler beams. This greatly enhances the practicality of the system by, for instance, eliminating the need for a quantum memory to store the idler. We perform a proof-of-principle experiment in the microwave regime, directly comparing the performance of a two-mode squeezed source to an ideal classical noise source that saturates the classical bound for correlation. We find that, even in the presence of significant added noise and loss, the quantum source outperforms the classical source by as much as an order of magnitude.
\end{abstract}

\date{\today}

\maketitle



Quantum illumination has recently gained attention as a possible avenue to improve the sensitivity of radar and other target detection technologies.\cite{Lloyd:2008hb,Tan:2008ht}  The approach takes advantage of strong signal correlations that can be created in electromagnetic beams using quantum processes. These quantum correlations, a form of entanglement, can be stronger than anything allowed by classical physics giving a ``quantum advantage" to the detection process. A number of proposals exist to use these correlations in a wide range of quantum sensing applications with the goal of making precision measurements beyond the standard quantum limit.\cite{Lanzagorta:2011ej,Degen:2017kw} Most of these applications require that the entire sensor system be low-noise and have negligible loss in order to maintain entanglement. Notably, quantum illumination seems to be very robust to the presence of background noise and loss, suggesting that it may have broader practical applications.

In this Letter, we present measurements demonstrating the potential of a novel quantum illumination protocol that implements a form of noise radar. Noise radar has been studied in the classical regime because of, among other reasons, the inherent difficulty in detecting the noisy probe beam against the ambient thermal background noise.\cite{Kulpa2013, Narayanan2016}  As discussed below, our protocol relaxes a challenging requirement of existing protocols, namely, joint measurement. This greatly increases the practicality of our scheme compared to others. In a proof-of-principle experiment, we use the protocol to demonstrate a quantum enhancement in the detected signal-to-noise ratio of an order of magnitude when comparing the performance of an entangled-photon source to an ideal classical noise source that saturates the classical bound for correlation.

At the heart of quantum illumination (QI) is a nonlinear quantum process known as parametric downconversion (PDC). In PDC, a strong pump beam with a high frequency, $f_p$, is incident on a nonlinear medium, resulting in the production of two lower frequency beams, commonly referred to as the signal and idler, such that the frequencies of the produced beams, $f_s$ and $f_i$, satisfy the relation  $f_p = f_s + f_i$. Roughly, we can think of one pump photon being split to produce one signal and one idler photon, with the frequency condition ensuring conservation of energy. This process produces a type of entanglement know as two-mode squeezing (TMS).\cite{Caves:1985dn}  TMS manifests as strong correlations between the electric fields of the signal and idler. An interesting and important characteristic of a TMS state is that, if one measures the signal or idler beams individually, they appear simply as thermal noise.  It is only when both beams are measured and compared to each other that the special quantum properties can be inferred. The apparent thermal nature of the signal and idler motivates us to make the comparison to noise radar.

The basic premise of using QI for long-range sensing is that the sensing transceiver would contain a PDC source.  One of the generated beams, say the idler, would be maintained by the transceiver, giving a record to later be compared to the signal. The signal beam would then be transmitted into the detection region.  On the receiver side, a QI sensing system would operate in a similar fashion to a noise radar system.\cite{Kulpa2013, Narayanan2016} Instead of simply measuring the power or amplitude of a returning signal, the receiver correlates the detected signal with the idler.  Only if a correlation is detected, would we infer the presence of an object that had reflected some part of the signal beam.  Even if there is significant background noise, as there always is in the microwave regime, that noise would be uncorrelated with the idler and would be rejected.  In the limit of very weak signals, quantum mechanics tells us that correlations between the signal and idler beams in a TMS can be significantly stronger than those allowed for classical beams.  Because of this, theory predicts that the detection efficiency of a QI system could exceed the efficiency of a classical system using the same power. Recent theoretical work suggests that the enhancement can even persist in the high-power regime.\cite{Wilde:2017fm}

Recent experiments have demonstrated the basic principle of quantum illumination using PDC at optical frequencies.\cite{Lopaeva:2013ik, Zhang:2013ha,England:2018wg} These are important results, but conventional radar systems typically use microwave frequencies.  It is therefore of interest to demonstrate QI in the microwave regime. One proposal \cite{Barzanjeh:2015id} has suggested using an optomechanical optical-to-microwave transducer to accomplish this.  In this Letter, we will study the direct production of quantum microwaves using a superconducting circuit.

In the QI protocols discussed in the above references, it is assumed that the idler beam is maintained \textit{in vivo}, for instance, in a lossless delay line or more sophisticated quantum memory.  The returning signal beam is then combined with the idler beam and a joint measurement, such as an interference measurement, is performed.  While, in theory, this scheme offers the ultimate quantum performance, it creates a number of problems in practice.  Most importantly, it essentially requires foreknowledge of the range of the object to be sensed, as the idler and signal must be combined at the appropriate time or no correlation will be observed.

\begin{figure}
\centering
\includegraphics[width=0.8\linewidth]{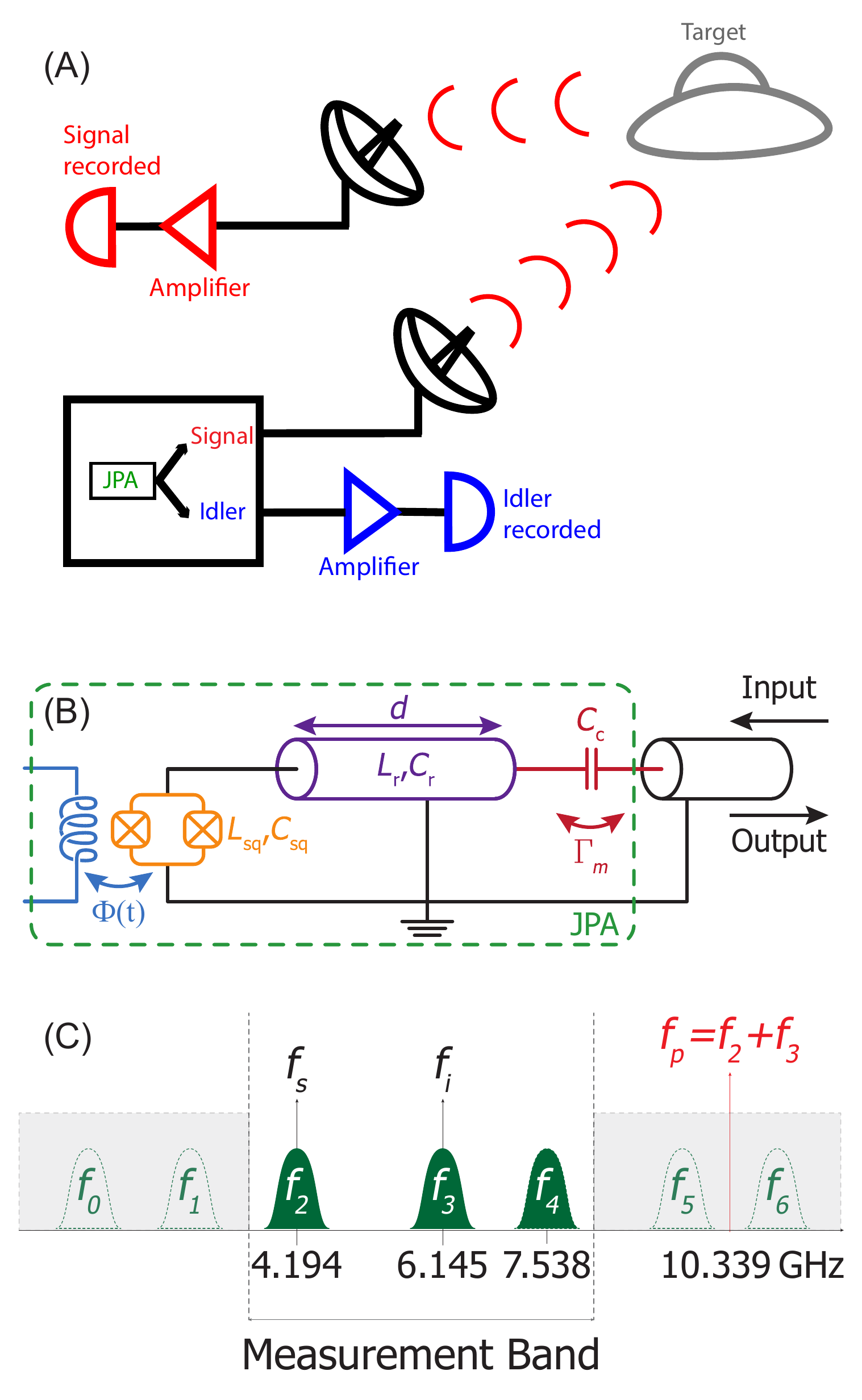}
\caption{\label{CartoonSetup} (A) Cartoon of the proposed quantum-enhanced noise radar protocol.  Amplifiers are used to measure both quadratures of the signal and idler. The signal and idler need not be measured simultaneously.  (B) Cartoon of the Josephson parametric amplifier used as a microwave parametric downconversion source. A micrograph of the device can be found in Chang \textit{et al}.\cite{W:2018ic} (C) The frequency modes of the JPA.}
\end{figure}

Here we propose a more practical protocol (see Fig.~\ref{CartoonSetup}) that we call quantum-enhanced noise radar (QENR). We measure the idler beam immediately, converting it into a classical record that can be copied and stored.  The returning signal is then measured and also converted to a classical record. The signal and idler records can then be correlated digitally over arbitrarily long time delays.  This greatly simplifies the operation of the proposed system, making it much more flexible and capable. Depending on the details of the measurement protocol, the cost of giving up the joint quantum measurement may be a reduction in the theoretical signal-to-noise ratio resulting from additional amplifier noise. However, in practical target-detection scenarios, this additional noise would be small compared to ambient noise and loss.  Regardless, as we show below, this protocol gives a significant quantum enhancement over a classical noise radar.


As a quantum microwave source, we  use a nondegenerate Josephson parametric amplifier (JPA).\cite{Yurke:1989we,CastellanosBeltran:2007ji,Yamamoto:2008cr,Wilson:2011ir,Flurin:2012hq,Simoen:2015by,W:2018ic} JPAs have been studied extensively in recent years because they work as ultrasensitive microwave amplifiers.  Conveniently for our purposes, their mechanism for amplification is PDC.  In the presence of an explicit input signal, PDC leads to amplification of this input. With no input signal, however, spontaneous PDC produces the TMS state needed for QI.  Alternatively, we can think of this spontaneous process as parametric amplification of quantum vacuum fluctuations.

Our JPAs are on-chip microwave circuits made from superconducting aluminum. Central to the JPA is a quarter-wavelength coplanar waveguide (CPW) resonator (see Fig.~\ref{CartoonSetup}). At one end, the CPW cavity is coupled to the input/output line through a small capacitance. At the other end, the cavity is terminated by a superconducting quantum interference device (SQUID), which acts as a parametric inductance that can be tuned by an external magnetic field.  For a fixed inductance, the cavity has a series of resonance frequencies at $f_0$, $f_1 \approx 3f_0$, $f_2 \approx 5f_0$, etc. By tuning the inductance, we can shift the fundamental resonance frequency $f_0$.  For this application, we chose the fundamental frequency to be $f_0 \approx 1$ GHz, giving us modes spaced by approximately 2 GHz.  This allows us to have 2-3 modes in our typical measurement band of 4-8 GHz.

If we pump the flux through the SQUID at the sum of two of the resonance frequencies, i.e.  $f_p = f_2 +f_3$, we then get downconversion into the modes $f_2$ and $f_3$. These are then the signal and idler of our TMS state (see Fig.~\ref{CartoonSetup}).  Ideally, both the output power of the modes, $P_i$, and the covariance between them, $C$, depend on the squeezing parameter $r$, which is generally a monotonically increasing function of pump power.  For an ideal two-mode squeezer, we can show that $P_i = \cosh(2r)$ and $C = \sinh(2r)$, with the powers expressed in units of photon number.\cite{Adesso:2005fw}

The quantum state of the microwave fields produced by the JPA can be fully characterized by measuring the covariance matrix of the corresponding in-phase, $I$, and quadrature, $Q$, voltages. This is a general property of so-called Gaussian states, which include the classical thermal and coherent states, as well as squeezed states. $I$ and $Q$ are common concepts in modern wireless and radar technology.  In the quantum realm, they are the canonically conjugate variables of the electromagnetic field, analogous to position and momentum in a mechanical system.  If we consider the signal and idler modes, labeled $s$ and $i$, we have a 4-by-4 covariance matrix:
\begin{align*}
\mathbf{V'}=
	\begin{pmatrix}
	I_s^2	&	I_sQ_s	&	I_sI_i	&	I_sQ_i\\
	Q_sI_s	&	Q_s^2	&	Q_sI_i	&	Q_sQ_i\\
	I_iI_s	&	I_iQ_s	&	I_i^2		&	I_iQ_i\\
	Q_iI_s	&	Q_iQ_s	&	Q_iI_i	&	Q_i^2\\
	\end{pmatrix}.
\end{align*}
In order to quantify quantum properties such as entanglement, the room-temperature covariance matrix $\mathbf{V'}$ must be calibrated and normalized in units of absolute photon number, yielding a scaled covariance matrix, $\mathbf{V}$. To characterize entanglement in $\mathbf{V}$, we use a common test known as the positive partial transpose (PPT) criterion.\cite{Simon:1994gb,Simon:2000fd,Adesso:2005fw} Using the PPT criterion, the degree of entanglement can be quantified by the minimum symplectic eigenvalue, $\widetilde{\nu}_{\text{min}}$, of the partial transpose of $\mathbf{V}$. The entanglement condition is $\widetilde{\nu}_{\text{min}}< 1$.  For details of the calibration procedure and entanglement test, see Chang \textit{et al}.\cite{W:2018ic}

We now characterize the performance of our QENR protocol in the presence of added noise and loss. For our classical illumination comparison, we will use the closest classical analog of our protocol, which is noise radar.  For our classical signal, we will use the ideal classical analog of our quantum signal.  By ideal, we mean that the correlations in the state saturate the classical bound.


We can produce this ideal classical signal in a simple way. We use a vector generator to generate band-limited Gaussian noise centered at $\approx1$ GHz. This noise is then mixed with a carrier at $\approx5$ GHz.  This produces two sidebands of noise that have the same correlation structure as our quantum source.  In the experiment, the frequencies are chosen such that the center frequencies of the sidebands exactly matched the frequencies of our quantum signal and idler.  In addition, the correlation structure of the quadratures is the same. At room temperature, we verify that the correlation between the classical signal and idler sidebands is 99\%, limited by small experimental imperfections.  To compare to our quantum source, we inject the classical beams into the cryostat and attenuate them down to the single-photon level. At low temperature, we reflect them off of the unpumped JPA, where it then enters the detection chain in exactly the same way as the quantum beams produced by the JPA.  We then process the classical and quantum beams in exactly the same way.
 
 \begin{figure}
\centering
\includegraphics[width=1\linewidth]{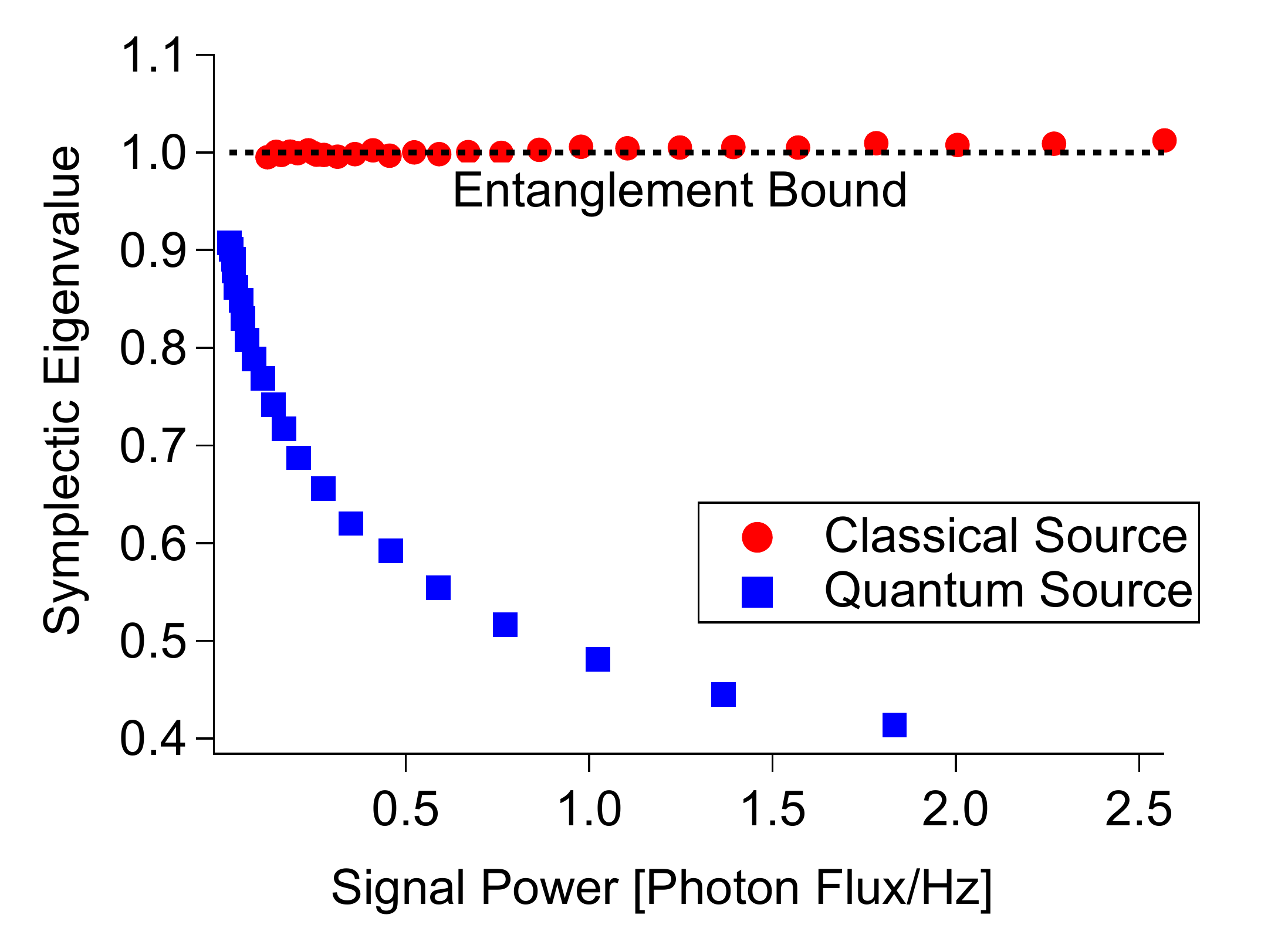}	
\caption{\label{Entanglement}Characterizing the ``quantumness" of our two sources as a function of output power. The minimum symplectic eigenvalue, $\widetilde{\nu}_{\text{min}}$, described in the text, characterizes entanglement between the signal and idler, with $\widetilde{\nu}_{\text{min}}<1$ implying that they are entangled.  }
\end{figure}
 
As mentioned in the introduction, one of the most promising aspects of QI is its resiliency to noise and loss.  While this proof-of-principle experiment takes place fully inside of a cryostat, we can simply use our amplifier noise and signal chain loss to simulate ambient noise and loss.  The noise and loss are therefore fixed.  However, we can still vary the signal-to-noise ratio of the system by changing the signal level, which is done in the quantum case by changing the pump power. We have calibrated the system noise temperature, which combines noise and loss into one figure of merit, to be $T_N\approx8$K. This corresponds to adding about 30 photons/Hz of noise, compared to our 0.1-1 photon/Hz signal levels. 

We measured the detected correlation between the signal and idler beams for both our classical and quantum sources. We studied this as a function of the raw SNR by varying the signal and idler power for fixed added noise.  In Fig.~\ref{Entanglement}, we characterize the ``quantumness" of our two sources.  For this plot, we characterize entanglement using $\widetilde{\nu}_{\text{min}}$.  Note that to calculate this quantity, we calibrate and subtract the system noise. Our results show that the classical source is always classical, although it saturates the correlation bound.  This is one sense in which our classical source is ideal. The quantum source always shows entanglement for the pump powers measured.  At higher pump powers, the degree of entanglement eventually saturates and then declines, indicating that some nonideal process is activated.

\begin{figure}
\centering
\includegraphics[width=1\linewidth]{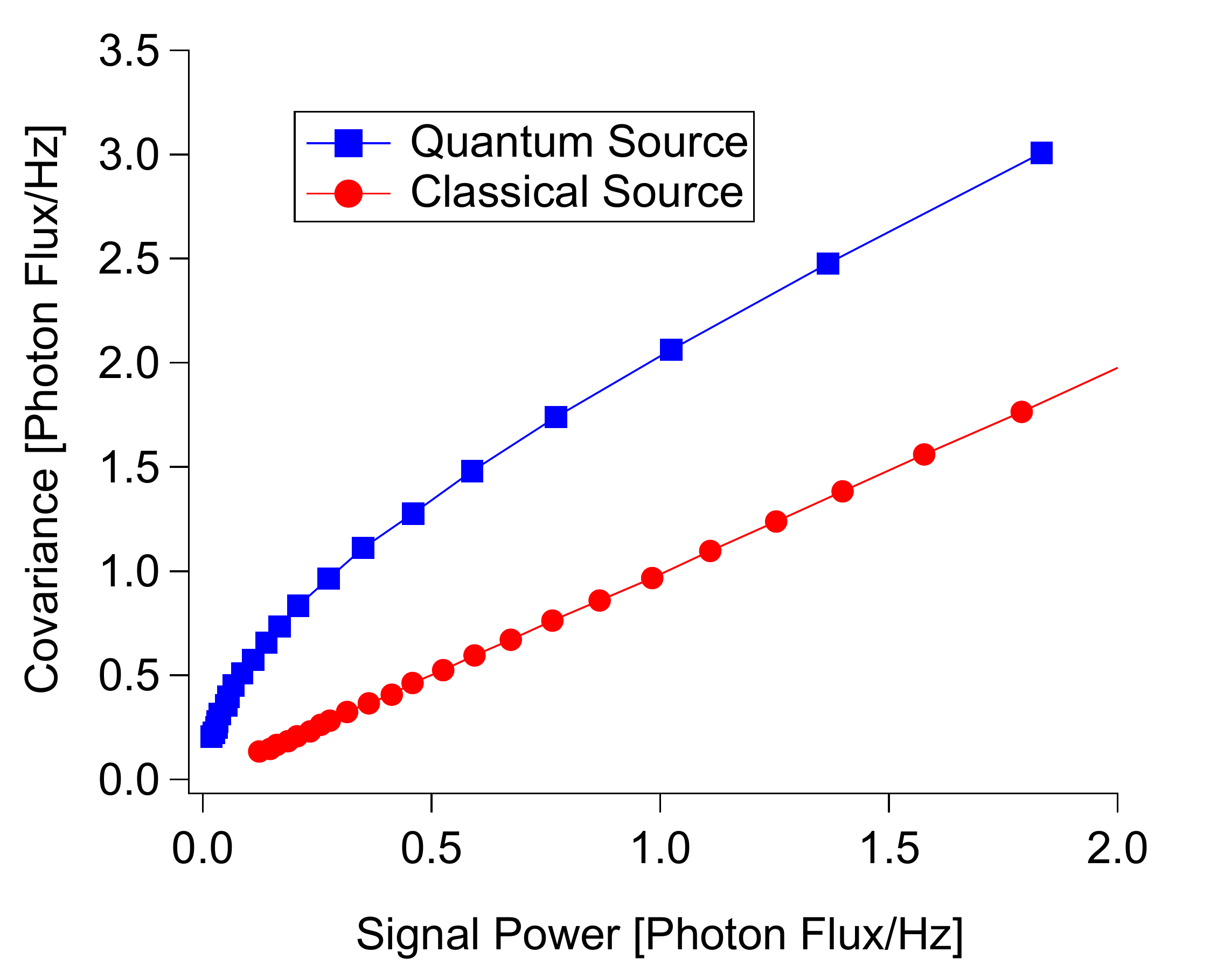}	
\caption{\label{RawCorr}Comparing the raw, detected covariance between the signal and idler for the quantum and classical source.  The bottom axis shows the source output power with system noise subtracted, $P_d$. The left axis shows the detected covariance with no subtraction.  At all powers, the covariance of the quantum source is higher than that of the classical source, illustrating the quantum enhancement.  }
\end{figure}

\begin{figure}
\centering
\includegraphics[width=1\linewidth]{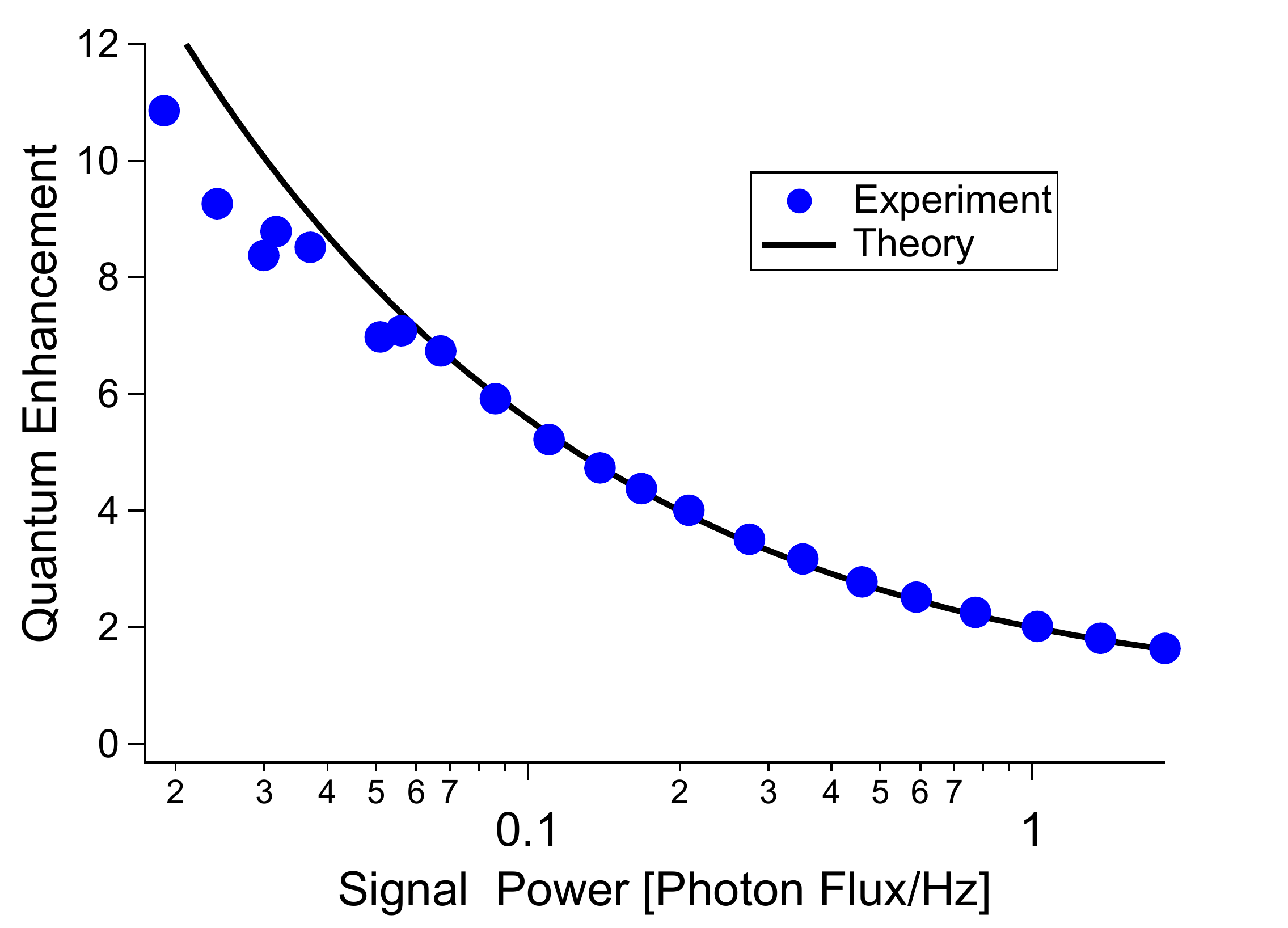}	
\caption{\label{QA}Quantum Enhancement.  We plot the ratio of the detected covariances for our quantum and classical sources as a function of $P_d$. The experimental data is fit to the simple expression in Eqn. 1, derived for an ideal parametric amplier. We see that at the lowest powers, the quantum enhancement can exceed a factor of ten. }
\end{figure}

To compare the performance of our classical and quantum source in the presence of noise and loss, we study the raw covariance including added noise, measured between the signal and idler modes.  These measurements directly simulate the detection principle of our proposed QENR protocol.  The noise and loss of our measurement chain simulates the noise and loss suffered by the signal beam propagating to the target and back. The raw covariance between the signal and idler channels is the ``detection signal" of our system. That is, the detection of a finite covariance indicates the presence of a target. The results are shown in Fig.~\ref{RawCorr}.  For the classical source, the detected covariance is 99\% of $P_d$, the measured source output power after subtracting the background noise.  This reproduces the correlation generated at room temperature.  Again, this illustrates that this is an ideal classical source.  Despite this, we see that the raw covariance of the quantum source is always higher, clearly indicating the quantum enhancement of our system.  This also indicates that, while the enhancement asymptotically vanishes with increasing power, it is nonetheless present. The same cannot be said for other quantum-enhanced measurements, such as those relying on N00N states.\cite{Lanzagorta:2011ej}

To make the enhancement more evident, in Fig.~\ref{QA}, we plot the ratio of the detected covariances for the quantum and classical source, which we will call the ``quantum enhancement." We see that for the lowest powers, the measured enhancement can exceed a factor of ten. 

Assuming that the device operates as an ideal JPA, we can derive a simple expression for the quantum enhancement, $E_Q$, namely,
\begin{equation}
E_Q = \sqrt{1 + 2\frac{P_0}{P_d}}
\end{equation}
where $P_0$ is a constant scale factor which includes the gain of the measurement system, $G$.  Figure 4 includes a fit of the data to this simple theoretical expression, with the single fitting parameter being $G$.  The extracted value of $G=61.1$ dB agrees well with the results of our independent calibration. We see that the achieved result is very close to the ideal.  

We note, however, that our noise radar analog may not be the optimal classical detection scheme. Still, the enhancement demonstrated has practical implications for certain applications. That is, since the QI source and noise radar beam look like thermal noise, they are difficult to distinguish from the ambient thermal background, even if they are absorbed by a detector.  In contrast, the man-made nature of, for instance, a coherent probe belies the presence of the radar system.  In applications where it is desirable to avoid detection of the radar system, the quantum enhancement demonstrated here would allow for the probe power to be reduced, further improving the undetectability of the system.


In conclusion, we have developed a source of entangled microwave photons that is suitable for testing quantum illumination protocols.  We have tested the performance of the quantum source and compared it to an ideal classical noise radar source. Even in the presence of significant added noise, we show that our QENR detection scheme can give a signal-to-noise ratio improvement exceeding a factor of ten compared to the analogous classical detection scheme.   This is clearly a promising first result.

The authors wish to thank B. Plourde, J.J. Nelson and M. Hutchings at Syracuse University for invaluable help in junction fabrication. The authors would  like to acknowledge D. Loung for helpful discussions and M. Simoen for help with illustrations. CMW, CWSC,  and AMV acknowledge NSERC of Canada; the Canadian Foundation for Innovation; the Ontario Ministry of Research, Innovation and Science; Canada First Research Excellence Fund (CFREF); Industry Canada; NRC of Canada; DRDC; and the CMC for financial support. JB acknowledges separate support from CFREF, NRC and DRDC.

\bibliography{QuantumIllumination.bib}

\end{document}